\documentstyle{mn2e}
\input{psfig}
\def\beq{\begin{equation}}
\def\eeq{\end{equation}}
\def\barray{\begin{eqnarray}}
\def\earray{\end{eqnarray}}

\newcommand{\etal}{{et al.~}}

\newcommand{\kmsmpc}{\>{\rm km}\,{\rm s}^{-1}\,{\rm Mpc}^{-1}}

\newcommand{\Msun}{\>{\rm M_{\odot}}}
\newcommand{\Lsun}{\>{\rm L_{\odot}}}

\def\gtsima{$\; \buildrel > \over \sim \;$}
\def\ltsima{$\; \buildrel < \over \sim \;$}
\def\prosima{$\; \buildrel \propto \over \sim \;$}
\def\gsim{\lower.7ex\hbox{\gtsima}}
\def\lsim{\lower.7ex\hbox{\ltsima}}
\def\simgt{\lower.7ex\hbox{\gtsima}}
\def\simlt{\lower.7ex\hbox{\ltsima}}
\def\simpr{\lower.7ex\hbox{\prosima}}
\def\la{\lsim}
\def\ga{\gsim}
\def\lta{\la}
\def\gta{\ga}

\newcommand{\apj}{ApJ}
\newcommand{\apjs}{ApJS}
\newcommand{\aj}{AJ}
\newcommand{\mnras}{MNRAS}

\newcommand{\nat}{Nature}

\newdimen\hssize
\newdimen\hdsize
\def\calC{{\cal C}}

\begin{document}

%%%%%%%%%%%%%%%%%%%%%%%%%%%%%%%%%%%%%%%%%%%%%%%%%%%%%%%%%%%%%%%%%%%%%%%%%%

\title[Properties of Galaxy Groups in the SDSS]  
      {Properties of Galaxy Groups in the SDSS: \\ 
       II.-- AGN Feedback and Star Formation Truncation}
\author[S.M. Weinmann et al.]          
       {\parbox[t]{\textwidth}{
        Simone M. Weinmann,$^{1}$\thanks{E-mail:weinmann@physik.unizh.ch}  
        Frank C. van den Bosch$^{2}$, Xiaohu Yang$^{3}$, H.J. Mo$^{2}$,\\   
        Darren J. Croton$^{5}$, Ben Moore$^{1}$}\\  
        \vspace*{3pt} \\
       $^1$Institute for  Theoretical  Physics, University  of
           Zurich,  CH-8057, Zurich,  Switzerland\\  
       $^2$Max-Planck-Institute for Astronomy, K\"onigstuhl 17,  
           D-69117 Heidelberg, Germany\\ 
       $^3$Shanghai Astronomical Observatory; the Partner Group of
           MPA, Nandan  Road 80, Shanghai  200030, China\\ 
       $^4$Department  of Astronomy,  University of Massachusetts, 
          710 North Pleasant Street, Amherst MA 01003-9305, USA\\ 
       $^5$Department of Astronomy, University of California at
           Berkeley, Mail Code 3411, Berkeley, CA 94720 USA}

%%%%%%%%%%%%%%%%%%%%%%%%%%%%%%%%%%%%%%%%%%%%%%%%%%%%%%%%%%%%%%%%%%%%%%%%%%

\date{}

\pubyear{2006}

\maketitle

\label{firstpage}

%%%%%%%%%%%%%%%%%%%%%%%%%%%%%%%%%%%%%%%%%%%%%%%%%%%%%%%%%%%%%%%%%%%%%%%%%%

\begin{abstract}
  Successfully  reproducing  the galaxy  luminosity  function and  the
  bimodality in the galaxy  distribution requires a mechanism that can
  truncate star formation in massive haloes.  Current models of galaxy
  formation  consider two  such truncation  mechanisms: strangulation,
  which  acts   on  satellite   galaxies,  and  AGN   feedback,  which
  predominantly affects  central galaxies.  The  efficiencies of these
  processes set the blue fraction of galaxies, $f_{\rm blue}(L,M)$, as
  function of  galaxy luminosity,  $L$, and halo  mass, $M$.   In this
  paper  we use  a galaxy  group  catalogue extracted  from the  Sloan
  Digital  Sky Survey  (SDSS)  to determine  $f_{\rm blue}(L,M)$.   To
  demonstrate the  potential power  of these data  as a  benchmark for
  galaxy   formation   models,  we   compare   the   results  to   the
  semi-analytical model  for galaxy formation of  Croton \etal (2006).
  Although this  model accurately fits the {\it  global} statistics of
  the  galaxy population,  as well  as  the shape  of the  conditional
  luminosity  function, there are  significant discrepancies  when the
  blue fraction  of galaxies as a  function of mass  and luminosity is
  compared between the observations and the model.  In particular, the
  model  predicts (i)  too many  faint satellite  galaxies  in massive
  haloes, (ii) a blue fraction of satellites that is much too low, and
  (iii)  a blue  fraction of  centrals that  is too  high and  with an
  inverted luminosity  dependence.  In the  same order, we  argue that
  these discrepancies owe to (i) the neglect of tidal stripping in the
  semi-analytical   model,  (ii)   the  oversimplified   treatment  of
  strangulation, and (iii) improper modeling of dust extinction and/or
  AGN feedback.  The data presented here will prove useful to test and
  calibrate  future models of  galaxy formation  and in  particular to
  discriminate between various models  for AGN feedback and other star
  formation  truncation   mechanisms.  

\end{abstract}

%%%%%%%%%%%%%%%%%%%%%%%%%%%%%%%%%%%%%%%%%%%%%%%%%%%%%%%%%%%%%%%%%%%%%%%%%%


%%%%%%%%%%%%%%%%%%%%%%%%%%%%%%%%%%%%%%%%%%%%%%%%%%%%%%%%%%%%%%%%%%%%%%%%%%

\section{Introduction}
\label{sec:intro}

One of  the most challenging outstanding problems  in galaxy formation
is to  explain the  detailed shape of  the galaxy  luminosity function
(hereafter LF).  In particular, the relatively shallow faint-end slope
and the  exponential cut-off at the  bright end of the  LF have proven
difficult to explain  (e.g., White \& Frenk 1991;  Benson \etal 2003). 
In  the traditional scenario  for galaxy  formation, it  is envisioned
that lower cooling efficiencies  in massive galaxies would explain the
exponential tail of the LF (Rees \& Ostriker 1977; Silk 1977; White \&
Rees 1978),  while supernova feedback  is typically invoked  to reduce
the star formation  efficiency in low mass haloes  (Larson 1974; White
\& Rees 1978; Dekel \& Silk  1986).  Although the latter can indeed be
tuned to  reproduce the faint-end slope  of the galaxy  LF, it worsens
the  problems at  the bright  end.  As  nicely demonstrated  by Benson
\etal  (2003), supernova  feedback causes  a drastic  increase  of the
amount of diffuse  hot gas that remains in larger  halos.  This gas is
able to cool onto the  central galaxies in these haloes, producing too
many  bright galaxies.   In  addition, this  causes  the bright  model
galaxies to have relatively young stellar populations, in disagreement
with  observations (e.g.,  Kauffmann  \& Charlot  1998; Heavens  \etal
2004;  Thomas \etal 2005).   What is  needed is  a mechanism  that can
truncate  the star  formation in  these massive,  central  galaxies at
relatively late times.

Star formation truncation  is also the main mechanism  that is thought
to underlie the bimodality  of galaxy properties. The local population
of galaxies consists roughly of  two types: red galaxies, which reveal
an early type morphology and which have very little or no ongoing star
formation,  and  blue  galaxies  with  active  star  formation  and  a
late-type morphology (e.g., Strateva  \etal 2001; Blanton \etal 2003a;
Kauffmann \etal 2003, 2004;  Baldry \etal 2004; Brinchmann \etal 2004;
Balogh  \etal 2004a,b).   Although  a non-negligible  fraction of  red
galaxies are clearly  edge-on disk galaxies that owe  their red colour
to  an enhanced  extinction, the  most pronounced  distinction between
`red  sequence' and  `blue sequence'  galaxies is  their  current star
formation rate.   Since only relatively small amounts  of ongoing star
formation tend to  make a galaxy appear `blue',  the colour bimodality
basically  reflects star formation  truncation: red  sequence galaxies
have their star formation  truncated, while blue sequence galaxies are
still forming stars today.

Semi-analytical  models  for galaxy  formation  consider  a number  of
mechanisms that can prevent, delay or truncate star formation.  In low
mass haloes one typically  invokes reionization and supernova feedback
in order  to suppress or truncate  star formation.  In  haloes with $M
\gta 10^{12} h^{-1} \Msun$, on which  we will focus in this paper, two
additional truncation  effects are considered.  The  first one, called
strangulation,  only affects satellite  galaxies.  As  soon as  a dark
matter halo is accreted by a larger halo, its central galaxy becomes a
satellite  galaxy.  It is  often assumed  that this  accretion process
causes the satellite  galaxy to be stripped of its  hot gas reservoir. 
Consequently, after  a delay time  in which the galaxy  consumes (part
of)  its cold  gas, star  formation  is truncated,  and the  satellite
galaxy becomes red (Larson,  Tinsley \& Caldwell 1980; Balogh, Navarro
\&  Morris 2000).   Ram-pressure  stripping (Gunn  \&  Gott 1972)  may
shorten the  time-delay, by also  stripping the satellite of  its cold
gas reservoir, but  since the time scale for  strangulation is already
relatively short, the addition of ram-pressure stripping does not have
a  large  impact.   Virtually  all semi-analytical  models  of  galaxy
formation, starting  with Kauffmann, White \&  Guiderdoni (1993), have
taken this strangulation  mechanism into account.  In fact,  it is the
main   mechanism   that   causes   red-sequence  model   galaxies   to
preferentially reside in overdense regions such as groups and clusters
of galaxies,  in good agreement with observations  (e.g.  Oemler 1974;
Dressler  1980; Hogg \etal  2004; Balogh  \etal 2004b;  Weinmann \etal
2006).

The  second  star  formation  truncation mechanism  that  operates  in
massive haloes  is feedback from  active galactic nuclei  (AGN), which
predominantly  affects  central   galaxies.   Although  the  potential
importance of  AGN feedback has  long been recognized (e.g.,  Tabor \&
Binney 1993; Ciotti \& Ostriker 1997), it only recently has been given
serious consideration  in galaxy  formation models.  This  has largely
been motivated by X-ray observations  which reveal that AGN can indeed
impact  the hot  IGM  of  galaxy clusters  (e.g.,  Fabian \etal  2003;
McNamara \etal 2005).  Numerous  recent studies have demonstrated that
the inclusion of  AGN feedback in galaxy formation  models can help to
explain the  bright end exponential  cut-off of the  galaxy luminosity
function (e.g.  Granato \etal  2004; Croton \etal 2006; Cattaneo \etal
2006a; Sijacki \&  Springel 2005; Bower \etal 2005)  and the fact that
the most massive galaxies contain the oldest stars (Croton \etal 2006;
Scannapieco, Silk \& Bouwens 2005; Bower \etal 2005).

Despite these successes, we are  still far from a proper understanding
of how  AGN feedback may establish  an equilibrium state  where it can
efficiently suppress star formation in  the centers of massive haloes. 
In  the studies mentioned  above, AGN  feedback is  typically modeled
using oversimplified, heuristic scaling relations, often based on very
different views of how AGN  feedback might operate.  In particular, it
is still unclear which mode of  AGN activity is most important for the
star formation truncation discussed  above; the merger induced 'quasar
mode' which leads  to an initial starburst followed  by a quenching of
star formation, or the 'radio  mode', which is caused by continual and
quiescent  accretion of hot  gas onto  the central  supermassive black
hole.   Hopkins \etal  (2006) have  suggested that  merger-induced AGN
activity (the  'quasar mode') is  responsible for the  transition from
blue, star forming  to red, passive galaxies.  Springel,  Di Matteo \&
Hernquist (2005a),  Menci \etal (2005)  and Kang, Jing \&  Silk (2005)
have shown that this `quasar  mode' feedback can indeed terminate star
formation and  expel the gas from  the center of the  galaxy, once the
supermassive black  holes become  sufficiently massive.  On  the other
hand, Croton \etal (2006),  Bower \etal (2005), Cattaneo \etal (2006a),
Nusser \etal  (2006) and Sijacki  \& Springel (2006) have  argued that
the  'radio  mode' of  AGN  activity  (or  AGN feedback  operating  in
quasi-hydrostatically  cooling  haloes)   is  the  main  mechanism  to
truncate star  formation in massive galaxies.  How  exactly this radio
mode feedback operates, however, is  still unclear, as is evident from
the  fact  that the  aforementioned  studies  all  use very  different
formulations.

All these  different AGN feedback models  mainly differ in  the way in
which the  feedback efficiency scales  with halo mass and  with galaxy
properties such as  black hole mass and gas  mass fraction.  Since AGN
feedback  causes star  formation truncation,  one way  to discriminate
between these various models  is therefore to investigate the relative
fractions of blue and red galaxies as function of halo mass and galaxy
properties.  Such a study will  also help to improve our understanding
of  strangulation,   the  star  formation   truncation  mechanism  for
satellites.   Although  strangulation is  typically  modeled as  being
independent of halo  mass, one might argue that the  ability of a host
halo  to strip  a subhalo  of  its hot  gas reservoir  depends on  the
presence and density of the hot corona of the host halo, which in turn
may well be mass dependent.  Again, knowledge of the fractions of blue
and red (satellite) galaxies as  function of halo mass should allow us
to discriminate between these different possibilities.

Nowadays,  with large  galaxy surveys,  such as  the Two  Degree Field
Galaxy Redshift Survey (Colless \etal  2001) and the Sloan Digital Sky
Survey (York \etal 2000), the number of galaxies is sufficiently large
that,  in principle,  one could  accurately measure  the red  and blue
fractions as  a function of various  variables. In this  paper, we use
our SDSS group catalogue, presented in Weinmann \etal (2006, hereafter
paper~I),  to  compute the  fractions  of  blue  and red  galaxies  as
function of  both halo mass  and galaxy luminosity.  To  emphasize the
potential constraining power  of these data we compare  our results to
the  semi-analytical galaxy  formation model  of Croton  \etal (2006),
which includes  both strangulation and `radio-mode'  AGN feedback.  We
show that  although this model  accurately fits the  galaxy luminosity
function, the color-bimodality, and many other {\it global} statistics
of the galaxy population, it  fails dramatically when it comes down to
the  blue  fraction  of  galaxies  as  a function  of  halo  mass  and
luminosity.   We  argue that  this  has its  origin  in  the way  that
strangulation and  AGN feedback have been incorporated  and we briefly
discuss possible  modifications.  The aim  of this paper,  however, is
not to  present a new,  improved model for star  formation truncation,
but merely  to present  observational constraints that  will hopefully
prove useful in discriminating between the various models.

This paper is organized as follows. Section~\ref{sec:sdssgc} describes
our  SDSS  group  catalogue,  which  we compare  to  a  similar  group
catalogue  extracted from  the semi-analytical  model of  Croton \etal
(2006) described in  Section~\ref{sec:sam}.  The actual  comparison is
presented   in  Section~\ref{sec:eco},   while  Section~\ref{sec:disc}
discusses the  possible implications  for galaxy formation  models. We
summarize our results in Section~\ref{sec:concl}.

\section{The SDSS Group Catalogue}
\label{sec:sdssgc}

\subsection{Data}
\label{sec:data}

The Sloan Digital Sky Survey (SDSS;  York \etal 2000) is a joint, five
passband ($u$,  $g$, $r$, $i$  and $z$) imaging  and medium-resolution
($R \sim 1800$) spectroscopic survey.   In this paper, we focus on the
subset of  galaxies that  are in the  New York  University Value-Added
Galaxy Catalogue (NYU-VAGC) based on  the SDSS Data Release 2 (Blanton
\etal 2005).  This NYU-VAGC  is based on an independent, significantly
improved data  reduction.  From this catalogue we  select all galaxies
with  an extinction corrected  apparent magnitude  brighter than  $r =
17.77$, with redshifts in the range  $0.01 \leq z \leq 0.20$, and with
a redshift completeness  $\calC > 0.7$.  This leaves  a grand total of
$184,425$ galaxies. In  what follows, we use $M_r$  and $^{0.1}M_r$ to
indicate the absolute magnitude  in the $r$-band, k-corrected to $z=0$
and $z=0.1$,  respectively. All k-corrections  are based on  the model
described in Blanton \etal (2003b).

We  split  the galaxies  into  `red'  and  `blue' subsamples  using  a
magnitude dependent cut, which roughly follows the observed bimodality
scale in the colour-magnitude relation:
\begin{equation}
\label{bimodal}
^{0.1}(g-r)_{\rm cut} = 0.7 - 0.032 \left[ ^{0.1}M_r + 16.5 \right]
\end{equation}
(cf. paper~I). In  what follows, we refer to  galaxies that are redder
and bluer  than $^{0.1}(g-r)_{\rm cut}$ as `red'  and `blue' galaxies,
respectively.

\subsection{The group-finding algorithm}
\label{sec:algorithm}

Our working definition  of a galaxy group is  the ensemble of galaxies
that reside in the same  dark matter parent halo; galaxies that reside
in subhaloes  are considered  to be group  members that belong  to the
parent  halo  in which  the  subhalo is  located.   We  have used  the
halo-based  group finder  developed  by Yang  \etal (2005a,  hereafter
YMBJ) in  order to  group the galaxies  in the above  mentioned galaxy
catalogue. This  particular group finder  has been optimized  to group
galaxies  according to  their common  dark matter  halo, and  has been
thoroughly tested  with mock galaxy  redshift surveys.  In  brief, the
method  works   as  follows.   First,  potential   group  centers  are
identified using a Friends-Of-Friends  (FOF) algorithm or an isolation
criterion.   Next, the total  group luminosity  is estimated  which is
converted  into  an estimate  for  the  group  mass using  an  assumed
mass-to-light ratio. From this  mass estimate, the radius and velocity
dispersion of  the corresponding dark matter halo  are estimated using
the virial equations,  which in turn are used  to select group members
in redshift  space.  This method  is iterated until  group memberships
converge.   A more  detailed  description is  given  in Appendix~A  of
paper~I.

In YMBJ the performance of this  group finder has been tested in terms
of  completeness of  true  members and  contamination by  interlopers,
using detailed mock galaxy redshift surveys.  The average completeness
of  individual groups was  found to  be $\sim  90$ percent,  with only
$\sim  20$  percent  interlopers.   Furthermore, the  resulting  group
catalogue  is  insensitive to  the  initial  assumption regarding  the
mass-to-light ratios, and the group finder is more successful than the
conventional FOF method (e.g.,  Huchra \& Geller 1982; Ramella, Geller
\& Huchra 1989; Merch\'an \&  Zandivarez 2002; Eke \etal 2004; Berlind
\etal  2006) in associating  galaxies according  to their  common dark
matter haloes.

\subsection{Estimating group masses}
\label{sec:groupmass}

Following YMBJ  we use  the group luminosity  to assign masses  to our
groups. The motivation  behind this is that one  naturally expects the
group luminosity to be strongly correlated with halo mass (albeit with
a certain amount of scatter).  Since the group luminosity is dominated
by  the brightest  members, which  are exactly  the ones  that  can be
observed in a flux limited  survey like the SDSS, the determination of
the (total) group  luminosity is more robust than  that of the group's
velocity dispersion,  especially when the  number of group  members is
small (see Appendix~B in paper~I).

Clearly, because of  the flux limit of the  SDSS, two identical groups
observed at different redshifts will have a different $L_{\rm group}$,
defined as  the summed luminosity  of all its identified  members.  To
circumvent this bias we first  need to bring the group luminosities to
a  common scale.  A nearby  group  selected in  an apparent  magnitude
limited  survey should  contain all  of its  members down  to  a faint
luminosity.  We can therefore use these nearby groups to determine the
relation  between the  group luminosity  obtained using  only galaxies
above a bright luminosity limit and that obtained using galaxies above
a  fainter   luminosity  limit.    Assuming  that  this   relation  is
redshift-independent,  one can  correct the  luminosity of  a high-$z$
group,  where   only  the  brightest  members  are   observed,  to  an
empirically normalized luminosity scale.

As  common  luminosity  scale   we  use  $L_{19.5}$,  defined  as  the
luminosity of  all group  members brighter than  $^{0.1}M_r =  -19.5 +
5\log  h$.  To  calibrate  the relation  between  $L_{\rm group}$  and
$L_{19.5}$  we first  select  all  groups with  $z  \leq 0.09$,  which
corresponds to the redshift for which a galaxy with $^{0.1}M_r = -19.5
+ 5\log  h$ has an apparent  magnitude that is equal  to the magnitude
limit of the  survey.  For groups with $z > 0.09$  we use this `local'
calibration  between $L_{\rm  group}$ and  $L_{19.5}$ to  estimate the
latter.   Detailed   tests  have   shown  that  the   resulting  group
luminosities are  significantly more reliable than those  in which the
correction  for  missing members  is  based  on  the assumption  of  a
universal luminosity function (see YMBJ for details).

The final step is to obtain  an estimate of the group (halo) mass from
$L_{19.5}$.   This  is done  using  the  assumption  that there  is  a
one-to-one relation between $L_{19.5}$  and halo mass.  For each group
we determine  the number density of  all groups brighter  (in terms of
$L_{19.5}$)  than the  group in  consideration.  Using  the  halo mass
function  corresponding to a  $\Lambda$CDM concordance  cosmology with
$\Omega_m=0.3$,  $\Omega_{\Lambda}=0.7$, $h=H_0/(100  \kmsmpc)  = 0.7$
and $\sigma_8=0.9$  we then find the  mass for which  the more massive
haloes have the same number density.  Although the masses thus derived
depend  on cosmology,  it  is straightforward  to  convert the  masses
derived here to any other cosmology. 

Finally we note  that not all groups can have a  halo mass assigned to
them.  First of all, the  mass estimator described above does not work
for groups in which all members  are fainter than $^{0.1}M_r = -19.5 +
5\log h$. Secondly, the combination  of $L_{19.5}$ and redshift may be
such that we  know that the halo catalogue  is incomplete, which means
that there is a significant number of groups at this redshift with the
same $L_{19.5}$ but for which the individual galaxies are too faint to
be detected.  Since our mass  assignment is based on the assumption of
completeness, any group beyond the completeness redshift corresponding
to its  $L_{19.5}$ is not assigned  a halo mass (see  Yang \etal 2005b
for details).

\subsection{The SDSS group catalogue}
\label{sec:catalogue}

Applying our group finder to  the sample of SDSS galaxies described in
Section~\ref{sec:data} yields a group catalogue of 53,229 systems with
an estimated  mass. These groups contain  a total of  92,315 galaxies. 
The  majority of  the groups  (37,216 systems)  contain only  a single
member, while there are 9220 binary systems, 3073 triplet systems, and
3720   systems  with   four   members  or   more   (see  paper~I   for
details)\footnote{This SDSS  group catalogue is  publicly available at
  http://www.astro.umass.edu/$^{\sim}$xhyang/Group.html}.    In   what
follows  we  refer  to the  brightest  galaxy  in  each group  as  the
`central' galaxy, while all others are termed `satellites'.
\begin{figure*}
\centerline{\psfig{figure=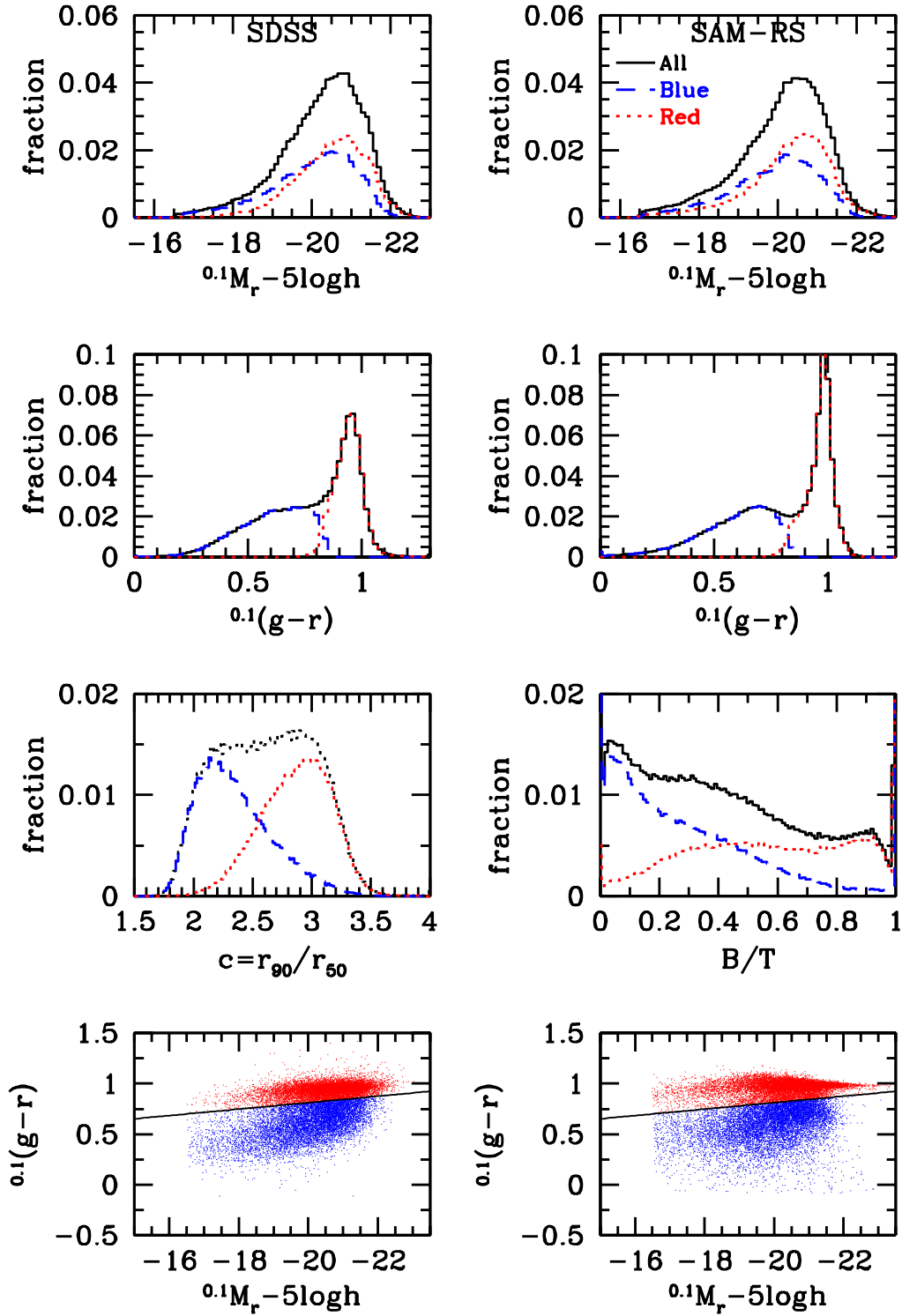,width=0.76\hdsize}}
\caption{A comparison  of global statistics of  the SAM-RS (right-hand
  panels) with  the SDSS (left-hand  panels). Panels in the  first and
  second row show histograms  of absolute $^{0.1}r$-band magnitude and
  $^{0.1}(g-r)$  colour for both  redshift surveys.  The contributions
  from blue and red galaxies are indicated by dashed and dotted lines,
  respectively.   Note the  good  agreement between  SDSS and  SAM-RS.
  Panels in the third row show histograms of morphological parameters.
  In  the  case  of  the   SDSS,  we  plot  the  distribution  of  the
  concentration  parameter $c$,  defined  as the  ratio  of the  radii
  containing 90 and 50 percent of  the petrosian flux.  In the case of
  the  SAM-RS, we  plot the  distribution of  the  bulge-to-total mass
  ratio $B/T$ instead. Although these can not be compared directly, in
  general a  more concentrated  galaxy (higher $c$)  will also  have a
  larger $B/T$.   Again, the contributions from red  and blue galaxies
  are  indicated.   Finally,  the  fourth  row  of  panels  shows  the
  colour-magnitude relations. The  solid line indicates the bimodality
  scale given  by equation~(\ref{bimodal}), which we use  to split the
  population of galaxies in red and blue subpopulations.}
\label{fig:1}
\end{figure*}

\section{The Semi-Analytic Model}
\label{sec:sam}

The semi-analytic model  (hereafter SAM) to which we  compare the SDSS
group  catalogue  discussed  above  is  based  on  an  output  of  the
Millennium  Run  $N$-body simulation  (Springel  \etal  2005b) and  is
described  in  detail in  Croton  \etal  (2006;  hereafter C06).   The
simulation  is based on  the cosmological  parameters $\Omega_m=0.25$,
$\Omega_{\Lambda}=0.75$,      $\Omega_b=0.045$,      $h=0.73$      and
$\sigma_{8}=0.9$  and has  a volume  of $0.125  h^{-3}{\rm  Gpc}^{3}$. 
Dark  matter  haloes are  identified  with  a  FOF group  finder,  and
subsequently  populated with  galaxies  following the  semi-analytical
model described in C06.
 
One of the  relative novelties of this SAM is  the inclusion of `radio
mode' feedback from AGN that lie at the center of a halo with a static
corona of hot gas.  As shown in C06, this feedback mode suppresses the
cooling flow in massive haloes at relatively late times, which in turn
yields luminosities, colours and  stellar ages for massive galaxies in
better agreement  with observations.  In particular,  the inclusion of
the radio-mode AGN feedback can explain the exponential cut-off at the
bright end  of the galaxy luminosity  function, and the  fact that the
most massive galaxies  are red and consist of  old stellar populations
(see De Lucia  \etal 2006 for details).  The  model also predicts star
formation histories,  cold gas  mass fractions and  metallicities that
are all in good agreement  with observations.  The model even predicts
a  Tully-Fisher zero-point  that  matches  the data,  as  long as  the
rotation velocity  of a disk galaxy  is equal to  the maximum circular
velocity of the  dark matter halo (however, see  Dutton \etal 2006 for
the unorthodox implications of this assumption).

Since  the halo  masses $M$  assigned to  our SDSS  galaxy  groups are
obtained by matching  the abundances to the halo  mass function, these
are  to  be  interpreted  as  the  masses  inside  a  radius  with  an
overdensity  of $180$.  As shown  by  Jenkins \etal  (2001), for  this
definition of halo mass the analytical halo mass function of Sheth, Mo
\& Tormen  (2001) used here  is in good  agreement with the  halo mass
function obtained from numerical  simulations.  The halo masses in the
C06 catalogue, however, are defined as the masses inside a radius with
a  mean density that  is $200$  times the  critical density,  which we
denote by $M_{200}$.  In order  to convert $M_{200}$ to $M$, we assume
that dark  matter haloes follow  a NFW density  distribution (Navarro,
Frenk \& White  1997).  Using the relation between  halo mass and halo
concentration of  Eke, Navarro \&  Steinmetz (2001), we find  that the
relation between $M$ and $M_{200}$ is well fit by
\begin{equation}
\label{mrat}
{M_{200} \over M} = 0.745 - 0.0006 \, 
\left[{\rm log}(M_{200})-7.0\right]^{2.45}\,.
\end{equation}

In what follows, we only consider the galaxies with $M_r \leq -16.72$,
which reflects the magnitude completeness limit of the SAM, yielding a
total of approximately 9 million model galaxies.  The main information
used in this  paper are the $r$ band and $g$  band magnitudes of these
galaxies,  and the  mass $M$  of the  halo in  which they  reside. For
comparison  with  the SDSS,  we  also  compute  the $g$  and  $r$-band
magnitudes k-corrected to $z=0.1$, using
\begin{equation}
^{0.1}g =g + 0.3113 + 0.4620 \, \left( g - r - 0.6102 \right)
\end{equation}
and
\begin{equation}
^{0.1}r =g - 0.4075 - 0.8577 \, \left( g - r - 0.6102 \right)
\end{equation}
(Blanton \etal 2003b)\footnote{These  filter transformations are taken
  from a manuscript in preparation  by Michael Blanton and Sam Roweis,
  available at http://cosmo.nyu.edu/blanton/kcorrect}.

As we will see below, despite  the AGN feedback the SAM still contains
a significant  number of  very bright and  blue galaxies that  are not
present in the SDSS.  As mentioned in C06, these are mainly ULIRG-type
starbursts for  which the dust  treatment of the model  is inadequate;
with  a more  proper dust  model, these  galaxies would  be  much more
extincted, making them  both fainter and redder. In  order to suppress
the impact of these galaxies  on our SAM-SDSS comparison we remove all
galaxies with $(g-r) <  0$ from both the SAM and the  SDSS. In the 
case of the SAM this  only affects 0.5 percent of all  galaxies, 
while in the case of the SDSS, this fraction is completely negligible.

We split the SAM model galaxies into `red' and `blue' subsamples using
the  same magnitude  dependent colour-cut  as for  the SDSS,  given by
equation~(\ref{bimodal}).    In  addition,  we   discriminate  between
`central' galaxies, defined as the galaxy in a halo that is closest to
the  halo center,  and `satellite'  galaxies.  This  differs  from the
definition used for the group  catalogues, where the central galaxy is
defined as the brightest group member.  However, $97.5$ percent of all
central galaxies  in the  SAM are also  the brightest galaxy  in their
halo,  virtually independent  of  halo mass.   We  have verified  that
defining central  galaxies in  the SAM as  the brightest  halo members
instead does not have a significant impact on any of our results.
\begin{figure}
\centerline{\psfig{figure=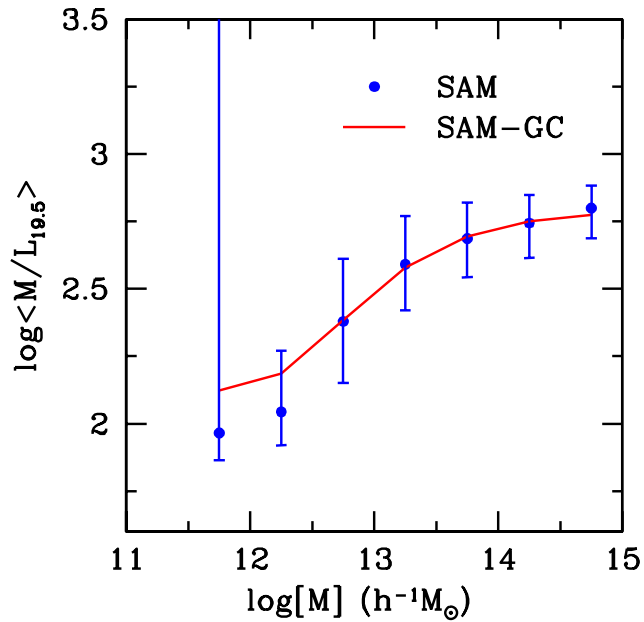,width=\hssize}}
\caption{The average mass-to-light ratio, $\langle M/L_{19.5} \rangle$
  as a  function of halo (group)  mass.  Here $L_{19.5}$  is the total
  luminosity of all galaxies in a halo with $^{0.1}M_r - 5 \log h \leq
  -19.5$.   Filled  circles with  errorbars  (indicating  the 68  $\%$
  confidence level)  show the results obtained from  the SAM directly,
  using the original halo masses  and halo membership.  The solid line
  shows  the $\langle  M/L_{19.5} \rangle$  as obtained  from  the SAM
  group catalogue, and is thus  based on the assigned group masses and
  the  assigned  group  memberships.   The  agreement  with  the  true
  $\langle M/L_{19.5} \rangle$ is excellent, indicating that our group
  finder allows  an accurate recovery of the  average relation between
  mass and light.}
\label{fig:ml}
\end{figure}

\subsection{Constructing a SAM redshift survey}
\label{sec:samsurvey}

In order to be able to compare  the SAM to the SDSS results we need to
mimic the construction of a galaxy redshift survey. We do so using the
following  steps.  First,  we construct  a large  virtual  universe by
replicating  the periodic simulation  box in  a stack  of $2  \times 2
\times 2$ boxes.  This is required in order to be able to probe out to
sufficiently high redshifts. Next we compute the redshift and apparent
magnitude of  each galaxy as seen  by a virtual observer  located in a
corner of this virtual universe (who can thus see $\pi/2$ steradian of
`sky').   We mimic  the selection  criteria of  the SDSS  discussed in
Section~\ref{sec:data} by only selecting those galaxies with $0.01 < z
< 0.2$  and with $r <  17.77$.  This leaves  us with a grand  total of
$428,013$ model galaxies.  In what follows, we refer to this sample as
the ``SAM redshift survey'' (hereafter SAM-RS).

Fig.~\ref{fig:1}  compares a  number of  statistics of  these galaxies
with those  from the SDSS. The  upper panels plot  the distribution of
absolute magnitudes,  $^{0.1}M_r$, and  their contribution due  to red
(dotted curves)  and blue (dashed  curves) galaxies. The  agreement is
very satisfactory, consistent  with the fact that the  SAM matches the
observed  luminosity functions (see  C06).  The  second row  of panels
indicate  the  color  distributions.   Once again,  the  agreement  is
reasonable, although the bimodality  in the SAM-RS seems somewhat more
pronounced  than in the  SDSS, with  a somewhat  narrower `red  peak'. 
Nevertheless, with blue fractions of 48 and 46 percent in the SDSS and
SAM-RS, respectively, the overall agreement is very satisfactory.

In  the third  row of  panels, we  `compare' two  different morphology
indicators.  For  the SDSS  galaxies we plot  the distribution  of the
concentration $c$, defined as the ratio between the radii that contain
90 and 50  percent of the Petrosian flux. For  the SAM model galaxies,
we  plot the  distribution of  the bulge-to-total  (B/T)  stellar mass
ratio instead. Typically, a galaxy with a large $B/T$ will also have a
high concentration parameter.  For both  $c$ and $B/T$ there is a very
significant overlap of red and  blue galaxies. Therefore, our split in
red  and   blue  galaxies  does   not  necessarily  correspond   to  a
morphological  split in  early and  late-type  galaxies, respectively,
even though both are clearly correlated (see Fig.~1 in paper~I).

Finally, the  lower panels of  Fig.~\ref{fig:1} show scatter  plots of
the  color-magnitude  relations. The  solid  line  corresponds to  the
bimodality  scale given by  equation~(\ref{bimodal}). Again,  there is
reasonable overall agreement between the SAM-RS and the SDSS, although
there  are  more  galaxies  with  very blue  colours  in  the  SAM-RS,
especially at  the bright end. Recall  that galaxies with  $(g-r) < 0$
have  already been excluded  from these  plots.  As  mentioned before,
these  bright, blue  galaxies would  appear significantly  fainter and
redder  with  proper  dust  modeling.  Another  apparent  discrepancy
concerns  the   red  sequence,  which   at  the  bright   end  appears
significantly tighter  for the  SAM than for  the SDSS, which  is also
apparent from the histograms in the second row of panels.

To   summarize,  despite   some  small   discrepancies,   the  global,
statistical properties of the SAM model galaxies are in good agreement
with the SDSS.  However, this does not mean that the SAM also predicts
the  correct statistics  {\it as  a function  of halo  mass}.  This is
clearly a much tighter constraint for  the model, and, as we argued in
Section~\ref{sec:intro},  may provide  useful  insights regarding  the
halo-mass dependence of various physical processes.
\begin{figure*}
\centerline{\psfig{figure=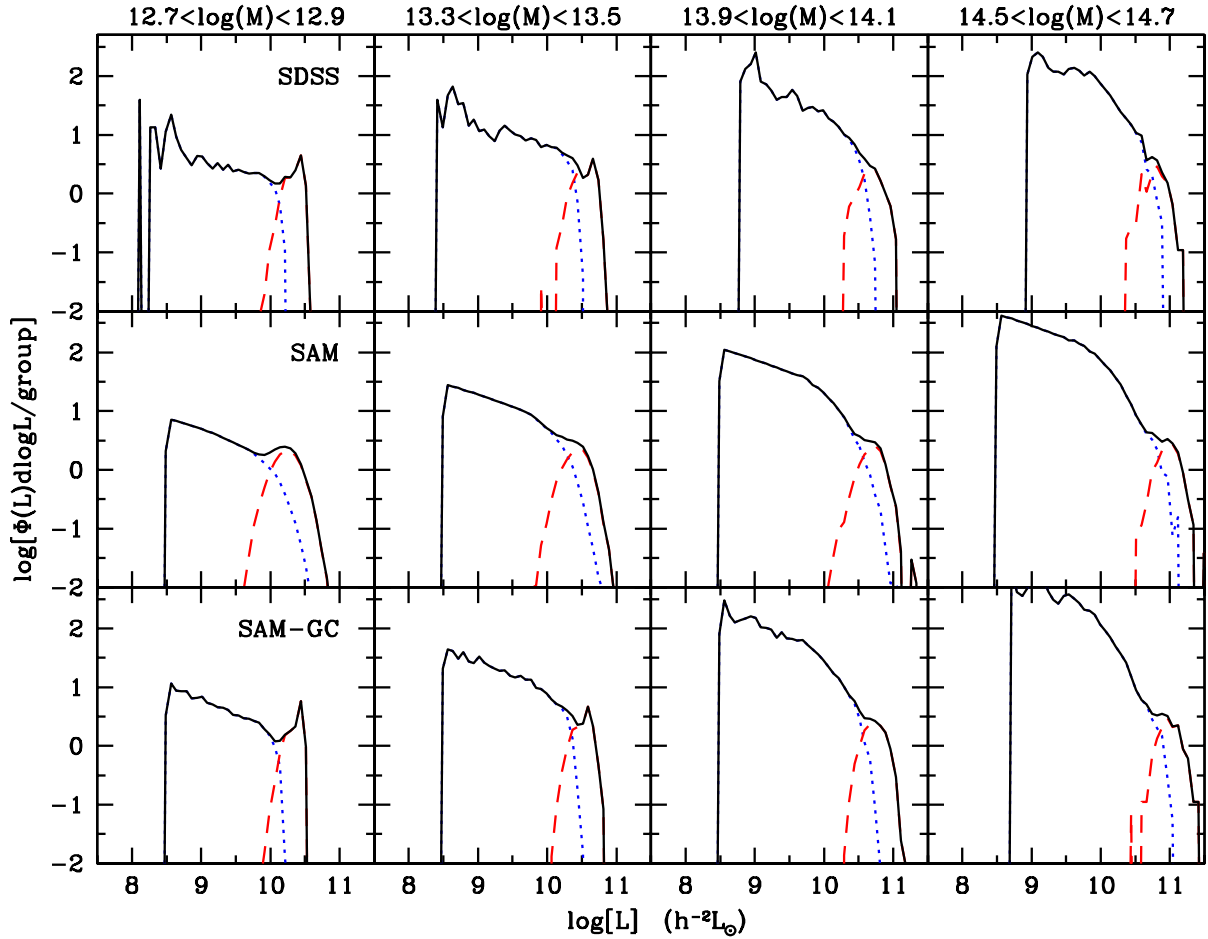,width=0.9\hdsize}}
\caption{The  conditional  luminosity  function obtained from the
  SDSS group catalogue (upper panel) from the SAM directly, using true
  halo masses  and true halo members  (middle panel) and  from the SAM
  group   catalogue,  using   assigned  group   masses   and  assigned
  memberships (lower  panels).  Results  are shown for  four different
  mass bins  (masses in  $h^{-1} \Msun$), as  indicated at the  top of
  each column.   Luminosities are  in the $^{0.1}r$-band.   The dotted
  (blue) line marks the  contribution from the satellite galaxies, the
  dashed (red) line the contribution from the central galaxies.}
\label{fig:clf_r}
\end{figure*}

\subsection{Construction of the SAM group catalogue}
\label{sec:samgc}

The main  goal of this  paper is to  compare the fractions of  red and
blue galaxies  as a function of  both halo mass and  luminosity in the
SAM with those  obtained from our SDSS group  catalogue.  In principle
we could do  so by comparing our SDSS group  results directly with the
SAM.  However, our group finder,  and in particular the algorithm used
to assign  the masses, is not  perfect.  Hence, it  is unclear whether
any  discrepancy between  SDSS  group catalogue  and  SAM reflects  an
artefact of the group finder, or whether there is a true difference in
the halo  occupation statistics  of SDSS and  SAM. To  circumvent this
problem, we use the SAM-RS  described above to construct a ``SAM group
catalogue'' (hereafter  SAM-GC), using  exactly the same  group finder
and  mass assignment  algorithm as  for  the SDSS.  By comparing  this
SAM-GC  with the  SDSS  group catalogue  we  significantly reduce  the
impact of  possible inaccuracies related  to the group  finder, making
the comparison more fair.

Application of our  group finder to the SAM-RS  described above yields
98,130 groups  with an  assigned mass, which  host a total  of 206,076
galaxies.  This  amounts to  an average of  $2.10$ members  per group,
which is significantly higher than for the SDSS, where the groups with
an assigned mass have on average $1.73$ members.  However, the SDSS is
not complete; in fact, the  average completeness of our SDSS sample is
$0.88$, which  largely explains the  difference in the mean  number of
members per group.  Another reason for this discrepancy is that, as we
will see, massive haloes in  the SAM contain too many faint satellites
compared to the SDSS.

The  fraction of  blue galaxies  in  the SAM-GC  is only  29 percent.  
Comparing this to  the fraction of 46 percent of  blue galaxies in the
SAM-RS  indicates  that  a  red  galaxy  is much  more  likely  to  be
associated with a group than a  blue galaxy. At first sight this seems
a  logical consequence of  the fact  that (i)  our group  catalogue is
limited to  relatively massive  haloes with $M  \gta 5  \times 10^{11}
h^{-1} \Msun$  and (ii) low mass  haloes are more likely  to host blue
galaxies, see e.g. paper~1. However, the application of the
group finder  to the SDSS only  reduces the fraction  of blue galaxies
from 48 percent to 41 percent. This reduction is much less severe than
for the SAM. This is the first indication that the SAM and SDSS do not
agree well when it comes down to details regarding the distribution of
red and blue galaxies (see Section~\ref{sec:fblum} below).

Since the  SAM contains the  full halo occupation information,  we can
check whether  our group finder  has assigned the correct  galaxies to
the same group, and whether the assigned mass is in agreement with the
true  halo  mass.   We have  performed  a  large  number of  tests  to
investigate how  well the group finder  allows us to  recover the true
relations between  galaxies and their dark matter  haloes.  Several of
these tests have been described in detail in Yang \etal (2005a,b), and
show that the  average occupation statistic of dark  matter haloes are
accurately recovered.  However, higher-order moments of the occupation
statistics, such as the scatter  around a mean relation, are typically
severely underestimated by  the group catalogue, due to  the fact that
we assume a one-to-one relation  between halo mass and halo luminosity
(with  zero scatter)  when  assigning  masses to  our  groups.  As  an
illustration of  the accuracy  of our group  finder, Fig.~\ref{fig:ml}
plots the average mass-to-light ratio of the dark matter haloes in the
SAM  (symbols with  errorbars).  Here  $M$  is the  halo mass  defined
according to equation~(\ref{mrat}), and $L$ is the total luminosity in
the $^{0.1}r$  band of all galaxies  in that halo with  $^{0.1}M_r - 5
\log h  \leq -19.5$.   The solid line  in Fig.~\ref{fig:ml}  shows the
average  mass-to-light ratios  obtained from  the SAM-GC,  which agree
extremely  well   with  the   true  $\langle  M/L   \rangle_M$.   This
demonstrates that our group  finder can accurately recover the average
relation between mass and light.  

In  what follows,  whenever we  present any  result obtained  from the
SAM-GC, we will also present  the same results extracted directly from
the SAM (using the true halo  masses and the true halo memberships). A
comparison among  these results safeguards  against potential problems
with the group finder.

\section{Galaxy Ecology}
\label{sec:eco}

\subsection{Conditional Luminosity Functions}
\label{sec:CLF}

We  start  our SAM-SDSS  comparison  by  focusing  on the  conditional
luminosity  function (CLF),  $\Phi(L  \vert M)$,  which specifies  the
average number of galaxies of luminosity  $L$ that reside in a halo of
mass $M$ (Yang,  Mo \& van den  Bosch 2003; van den Bosch,  Yang \& Mo
2003). The upper panels of Fig.~\ref{fig:clf_r} show the CLFs obtained
from the SDSS  group catalogue.  Results are shown  for four different
mass bins,  as indicated at  the top of  each column. Note  that these
masses are  the assigned group  masses.  The dotted (blue)  and dashed
(red) lines indicate the  contributions from the satellite and central
galaxies, respectively.  The distribution  of central galaxies is well
approximated  by a log-normal  distribution, consistent  with previous
findings  (Zheng \etal  2005; Yang  \etal  2005b). The  panels in  the
middle  row of  Fig.~\ref{fig:clf_r} show  the CLFs  obtained directly
from the SAM;  here the masses are the true halo  masses, and the true
halo members  are used to  construct the CLFs.  The  overall agreement
with  the  CLFs  extracted  from  the SDSS  group  catalogue  is  very
satisfactory, although  the width of  the CLF for central  galaxies is
significantly broader in the SAM than in the SDSS. To allow for a more
meaningful comparison,  the lower panels  of Fig.~\ref{fig:clf_r} plot
the CLFs obtained from the SAM-GC.   The first thing to notice is that
the width of  the CLFs of the central galaxies are  now in much better
agreement  with the  SDSS; apparently,  the group  finder artificially
`narrows'  the scatter  in  the  relation between  halo  mass and  the
luminosity of the  central galaxy.  This simply owes  to the fact that
we use the  group luminosity to determine the  group mass.  Other than
that, the  agreement between  the CLFs obtained  from the  SAM-GC, and
those extracted  directly from the  SAM is very good,  indicating that
our group finder allows an accurate recovery of the true $\Phi(L \vert
M)$ (see also tests in Yang \etal 2005b).

Except for the highest mass bin,  the CLFs extracted from the SDSS and
the  SAM group  catalogues  are  in good  agreement  with each  other,
indicating that the SAM not  only fits the galaxy luminosity function,
but it even does so as  function of halo mass. However, at the massive
end the  SAM predicts significantly more relatively  faint galaxies in
massive haloes then  observed. In the highest mass  bin shown, the SAM
overpredicts the  number of faint satellites  with $L =  3 \times 10^9
h^{-2} \Lsun$  by a  factor $\sim 2$.   Since these are  virtually all
red,  early-type galaxies  (see  below), this  suggests  that the  SAM
overpredicts the  number density of  faint, red galaxies.   Indeed, as
already shown in C06, the  SAM overpredicts the luminosity function of
red galaxies at  the faint end.  The analysis  here suggests that this
largely  owes to  an overabundance  of satellite  galaxies  in massive
haloes with $M \gta 3 \times 10^{14} h^{-1} \Msun$.
\begin{figure*}
\centerline{\psfig{figure=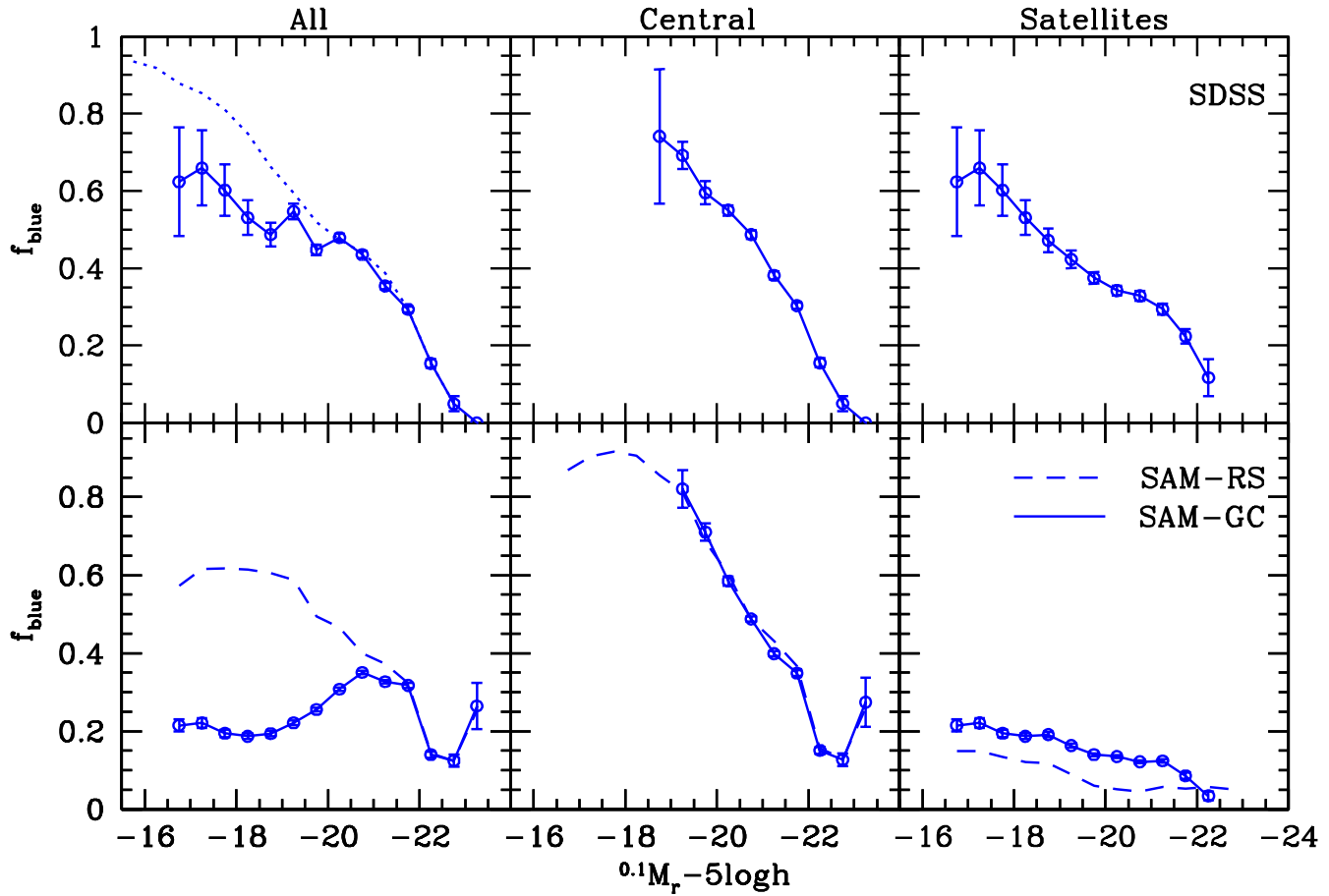,width=\hdsize}}
\caption{The luminosity dependence of blue galaxy fractions. From left
  to  right,  the panels  show  the  blue  fractions of  all  galaxies
  (centrals   plus  satellites),   central  galaxies,   and  satellite
  galaxies.   The top  panels show  the  results from  the SDSS  group
  catalogue (open circles with Poissonian errorbars).  The dotted line
  in the top-left panel shows the results obtained from the full SDSS,
  including those galaxies that were not assigned to a galaxy group by
  the group finder.   The bottom panels show the  results from the SAM
  of C06.   Results are shown for  both the SAM-RS  (dashed lines) and
  the   SAM-GC  (open   circles  with   Poissonian   errorbars).   See
  Section~\ref{sec:fblum} for a detailed discussion.}
\label{fig:3}
\end{figure*}

We emphasize that this discrepancy is not due to the fact that we have
ignored fiber  collisions in the SDSS. Since  the spectroscopic fibers
of  the  SDSS  have  a  minimum  angular  separation  of  $55''$,  the
spectroscopic  catalogue  suffers  from  an  incompleteness  on  small
angular scales. This  will impact on the multiplicity  function of the
groups in  our catalogue. However,  as shown by Berlind  \etal (2006),
the effect is relatively small, typically reducing the multiplicity of
groups  by $\sim  10$ percent,  which  is negligible  compared to  the
factor $2$ eluded to above.

\subsection{Blue fraction as Function of Luminosity}
\label{sec:fblum}

The dotted line in the  upper-left panel of Fig.~\ref{fig:3} shows the
fraction of blue  galaxies, $f_{\rm blue}$, in the  SDSS as a function
of luminosity. Here all $184,425$  galaxies in our SDSS sample defined
in Section~\ref{sec:data} are used. As  is well known, the fraction of
blue  galaxies  decreases   drastically  with  increasing  luminosity,
dropping from  $\sim 95$  percent at  $^{0.1}M_r - 5\log  h =  -16$ to
$\lta 5$ percent for galaxies with  $^{0.1}M_r - 5\log h < -22.5$. The
dashed line  in the lower-left panel  shows the blue  fraction for the
$428,013$ model  galaxies in the SAM-RS.  Although  this blue fraction
also reveals an overall decrease with increasing luminosity, there are
two marked differences with respect to the SDSS.  First of all, at the
bright end there  is a sudden upturn in  $f_{\rm blue}$; galaxies with
$^{0.1}M_r - 5\log h \simeq -23$  have a blue fraction of $\sim 26 \pm
6$ percent,  compared to  zero percent in  the SDSS (note  though that
there are  only 8 SDSS galaxies  in this luminosity  bin).  As already
mentioned in Section~\ref{sec:sam}, these  bright blue galaxies in the
SAM are ULIRGs for which  the dust modeling is inadequate (cf.  lower
panels  of Fig.~\ref{fig:1}).  The  second discrepancy  between SAM-RS
and SDSS is more important; at  the faint end the blue fraction in the
SAM-RS never exceeds 62 percent,  and is therewith much lower than the
blue  fraction  of  faint  SDSS  galaxies.  Consistent  with  what  we
inferred above  from the  CLFs, this indicates  that the  SAM severely
overpredicts the fraction of faint, red galaxies.

To  investigate   whether  this  mainly   concerns  central  galaxies,
satellite galaxies,  or both,  we now resort  to the  group catalogues
extracted  from the  SDSS and  SAM. The  open circles  with errorbars,
connected by solid lines, indicate the blue fractions of galaxies that
make  it into the  group catalogue.  Comparing these  for the  SDSS to
those obtained  from the full sample  (dotted lines), we  see that the
blue  fraction has become  somewhat lower  at the  faint end.   We can
understand this by  looking at the blue fractions  of central galaxies
(upper middle panel) and satellite galaxies (upper right panel).  This
shows that  both blue  fractions decrease with  increasing luminosity,
but  that  the  luminosity  dependence  is  more  pronounced  for  the
centrals. Since the galaxies that do  {\it not} make it into the group
catalogue are mainly isolated, and thus central, galaxies that live in
haloes below the  mass limit of the group  catalogue, these are mainly
blue.  This explains why the  fraction of faint blue galaxies is lower
in the group catalogue than in the full redshift catalogue.

In  the SAM,  the  group selection  also  causes a  drop  of the  blue
fraction of faint galaxies, but  of a much larger amplitude.  In fact,
the fraction of `group' galaxies with $-16 > M_{b_J} - 5 \log h > -18$
is $\sim 20$ percent in the  SAM, compared to $\sim 60$ percent in the
SDSS. The reason for this  discrepancy is largely due to the satellite
galaxies: as shown in the lower right-hand panel, the blue fraction of
satellite galaxies in the SAM is much too low, especially at the faint
end.   In  the case  of  the  central  galaxies (middle  panels),  the
agreement between SAM and SDSS is much better, especially for centrals
with $M_{b_J} - 5 \log h \simeq -21$. However, at the faint and bright
ends, the  SAM significantly overpredicts the blue  fractions by $\sim
15$ and $\sim 25$ percent, respectively.

The dashed lines in the  lower middle and lower right-hand panels show
the blue fractions of centrals and satellites of {\it all} galaxies in
the SAM-RS (including those that are not in the group catalogue). This
shows that the group finder very accurately recovers the blue fraction
of  central galaxies, but  slightly overpredicts  that of  satellites. 
This  owes to  the interlopers  (group  members that  do not  actually
belong to the same halo),  which tend to be isolated, central galaxies
in  low mass  haloes, and  which  are thus  preferentially blue.   The
contamination,  however,  is  sufficiently  small  that  it  does  not
significantly affect any of our results.

In  summary, although  the SAM  matches the  overall blue  fraction of
galaxies almost exactly  (see Section~\ref{sec:samsurvey}), when split
according to luminosity or according to centrals and satellites, there
are  very  significant differences  between  SAM  and  SDSS.  The  SAM
overpredicts the blue fraction of  central galaxies at both the bright
and the faint end  of the distribution, and dramatically underpredicts
the blue  fraction of (faint) satellite galaxies.   In particular, the
SAM predicts  that virtually all ($\gta 85$\%)  satellite galaxies are
red, whereas  the SDSS indicates  that the fraction of  red satellites
decreases from  $\sim 90$ percent at $^{0.1}M_r  = -22 + 5  \log h$ to
$\sim 40$  percent at  $^{0.1}M_r =  -17 + 5  \log h$.   This suggests
shortcomings for the star  formation truncation mechanisms in the SAM:
apparently the treatment of  strangulation is too efficient, while the
model   for    AGN   feedback    is   not   efficient    enough   (see
Section~\ref{sec:disc} for a more detailed discussion).
\begin{figure*}
\centerline{\psfig{figure=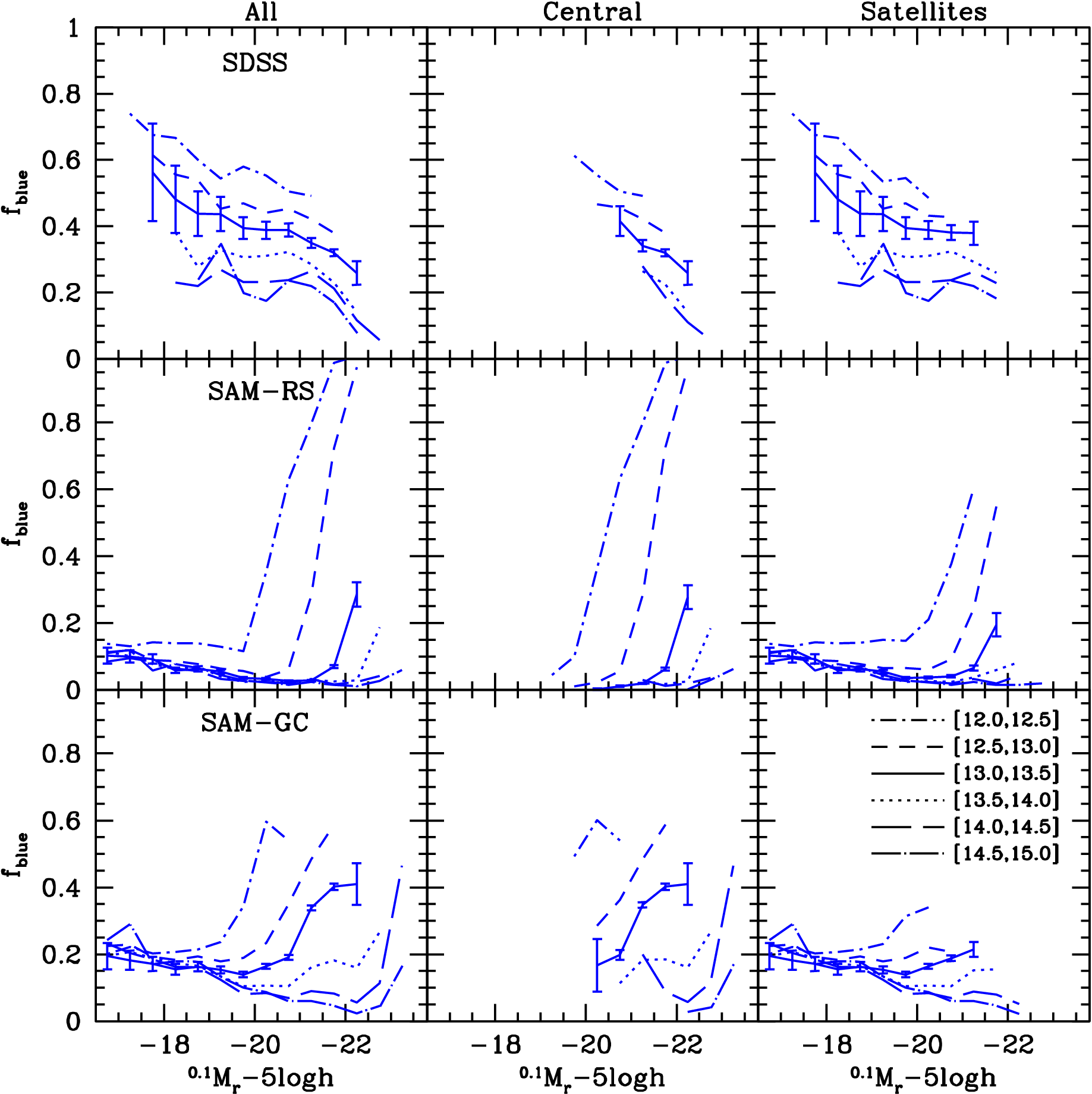,width=\hdsize}}
\caption{Blue galaxy fractions as a function of absolute magnitude in 
  the $^{0.1}r$-band.   Results are shown for  all galaxies (left-hand
  panels),  central galaxies  (middle panels)  and  satellite galaxies
  (right-hand panels), and for 6 different mass bins as indicated (the
  values   in   square   brackets   indicate  the   range   of   ${\rm
    log}(M/h^{-1}\Msun)$).  Results are only shown for mass-luminosity
  bins that  contain at  least 50 galaxies  in total, and  for clarity
  (Poissonian) errorbars are only shown for one mass bin.  From top to
  bottom,  results are  shown for  the SDSS  group catalogue,  the SAM
  redshift survey (SAM-RS), and the SAM group catalogue (SAM-GC). Note
  the poor agreement between SDSS  and SAM-GC, indicating that the SAM
  is not  correctly treating  the physics responsible  for determining
  whether a galaxy is red  or blue. See Section~\ref{sec:fbmass} for a
  detailed discussion.}
\label{fig:res}
\end{figure*}

\subsection{Blue Fraction as Function of Halo Mass}
\label{sec:fbmass}

We now  turn to the  mass dependence of  $f_{\rm blue}$. We  split the
SDSS  and  SAM group  catalogues  in  six  logarithmic mass  bins  and
determine  how the  blue fractions  in each  of these  bins  depend on
luminosity. For each  bin in mass and luminosity, the blue fraction is
defined as the  total number of blue galaxies in  that bin, divided by
the total number of galaxies in  that bin (i.e., we do not average the
blue fraction over individual groups or haloes).

The results for the SDSS group catalogue are shown in the upper panels
of Fig.~\ref{fig:res}  and are listed  in tabular form  in Appendix~A.
The upper left-hand panel shows  the result for all galaxies (centrals
plus satellites).  In each mass  bin, the blue fraction decreases with
increasing  luminosity,  but only  very  mildly.   In  fact, over  the
magnitude  range  $-19 \gta  ^{0.1}M_r  - 5  \log  h  \gta -21.5$  the
luminosity dependence  is remarkably weak,  for all six mass  bins. At
fixed luminosity, however, there is  a clear mass dependence, with the
blue  fraction decreasing with  increasing halo  mass. Over  the range
$10^{12}  h^{-1} \Msun  \lta M  \lta 10^{15}  h^{-1} \Msun$,  the blue
fraction  changes by  $\sim 30$  percent, at  all  luminosities.  This
indicates that the  colour of a galaxy is  more strongly determined by
the mass  of the halo in which  it resides than by  its own luminosity
(cf.  Yang \etal 2005b; paper~I). Consequently, the strong
luminosity dependence  of $f_{\rm  blue}$ seen in  Fig.~\ref{fig:3} is
mainly a reflection of the  fact that more luminous galaxies typically
reside  in   more  massive  haloes.   Note  that   some  earlier  work
(e.g. Balogh et al. 2004, Tanaka et al. 2004) has found no correlation
between galaxy  properties and halo velocity  dispersion.  However, as
shown in paper~I this is most likely due to the smaller
sample  size and the  fact that  using velocity  dispersion as  a mass
estimator tends to smear out the mass dependence.

The upper middle  panel shows the blue fractions  of central galaxies. 
Note  that at  a given  halo mass,  there is  only a  relatively small
dynamic  range  of luminosities  over  which  we  can measure  $f_{\rm
  blue}(L)$.   Nevertheless, at  a given  halo mass  there is  a clear
indication   that  the   blue  fraction   decreases   with  increasing
luminosity. At the same time,  at a given luminosity the blue fraction
also  decreases   with  increasing  halo  mass.   Finally,  the  upper
right-hand panel shows  the results for the satellite  galaxies in the
SDSS group catalogue, which look  similar to those for the full sample
of galaxies shown in the upper left-hand panel.

The middle  and lower  rows of panels  in Fig.~\ref{fig:res}  show the
results  obtained from the  SAM-RS and  SAM-GC, respectively.   In the
former  no group  finder has  been applied,  and the  mass  binning is
according to the  true halo masses of the  galaxies. A comparison with
the  results  obtained  from  the  SAM group  catalogue  is  therefore
indicative  of the  accuracy  with  which our  group  finder allows  a
recovery  of the  true underlying  $f_{\rm blue}(L,M)$.   A  number of
differences are  clearly apparent, which  mainly owe to the  impact of
interlopers  and, more importantly,  to errors  in the  assigned group
masses.   Since  the  true  $f_{\rm blue}(L,M)$  has  extremely  steep
gradients  in both  $L$ and  $M$, even  small errors  in any  of these
quantities can have a significant impact on the blue fraction obtained
from  the group  catalogue.  This  mainly  causes errors  in the  {\it
absolute} values of $f_{\rm  blue}(L,M)$.  However, the {\it relative}
relations of $f_{\rm  blue}(L \vert M)$ and $f_{\rm  blue}(M \vert L)$
are well recovered.   Note that the blue fraction  of central galaxies
with $^{0.1}M_r -  5 \log h =  -19.5$ is less than 10  percent for all
mass bins shown. Yet, as can  be seen from Fig.4, the blue fraction of
central galaxies in the group  catalogue with that luminosity is $\sim
80$  percent. This  indicates that  virtually all  blue  centrals with
$^{0.1}M_r - 5  \log h \leq -19.5$ reside in haloes  with $M < 10^{12}
h^{-1} \Msun$.  

Comparing the $f_{\rm blue}(L,M)$  obtained from the SAM-GC with those
obtained from  the SDSS, one  notices immediately that they  have very
little in  common. Probably the  most dramatic difference  between the
SDSS and the  SAM-GC concerns the blue fraction  of satellite galaxies
(shown in the panels on the right hand side), which is much too low in
the SAM-GC, especially for faint  galaxies and for low mass haloes.  A
comparison with the SAM-RS results shows that this discrepancy can not
be  attributed to  artefacts of  the group  finder.  In  addition, the
SAM-GC predicts  that the blue fraction of  central galaxies increases
with  increasing luminosity,  opposite to  what  is seen  in the  SDSS
(middle panels).  This effect is most severe for haloes with masses $M
< 10^{13}  h^{-1} \Msun$.   These two problems  conspire to  produce a
blue fraction of the  full galaxy population (centrals and satellites)
in  the  SAM-GC which  is  very different  from  the  SDSS (left  hand
panels), both quantitatively (mainly because  of the too low number of
blue satellites)  and qualititatively (mainly because  of the reversed
relation  between   luminosity  and  the  blue   fraction  of  central
galaxies).

We  are thus  led to  conclude that  although the  SAM  reproduces the
overall  blue fraction (when  integrated over  all galaxies),  when it
comes to  $f_{\rm blue}(L,M)$, there are  dramatic differences between
model and data.  Note that the  SAM of C06 fits the overall luminosity
function and even yields  CLFs, i.e., luminosity functions as function
of halo mass,  whose shapes are in remarkably  good agreement with the
SDSS data.  This suggests that one of the main problems for the SAM is
the  treatment of the  physics responsible  for determining  whether a
galaxy is  red or blue.   This includes the star  formation truncation
mechanisms,  such as  strangulation  and AGN  feedback,  but also  the
treatment of  dust extinction. In the  next section we  present a more
detailed discussion of the possible implications.

\section{Discussion}
\label{sec:disc}

The above  analysis of the fractions  of blue galaxies  as function of
halo mass and luminosity has  revealed several problems for the SAM of
C06.  In the following we discuss the possible implications for galaxy
formation models.

\subsection{Tidal Stripping}
\label{sec:tided}

The first  problem concerns the  abundance of satellite  galaxies.  As
shown  in Section~\ref{sec:CLF},  the SAM  overpredicts the  number of
faint satellite  galaxies in  massive haloes by  up to a  factor $\sim
2$. Most likely,  this indicates that the luminosity  evolution of the
satellites  is not  properly  accounted for.   Satellite galaxies  can
become fainter  due to star  formation truncation followed  by passive
evolution, or  due to  tidal stripping of  their stellar  mass.  Since
most of  the satellite galaxies in  the SAM are already  red, they can
not become much fainter than  they already are for their given stellar
mass.  In  other words, there is  little to gain  from adding physical
processes that may speed up the star formation truncation, such as ram
pressure stripping.  In fact, this  will only increase the fraction of
red satellites, which  is already much too large.   The most plausible
explanation for the overabundance of satellite galaxies is the neglect
of tidal stripping.  It is well known that the tides in massive haloes
can easily  strip satellite galaxies  and their dark matter  haloes of
large fractions of  their mass. This is supported  by the detection of
intracluster light  (e.g., Bernstein \etal 1995;  Gonzalez \etal 2000)
due to a diffuse population of intergalactic stars. Most likely, these
stars, which contribute  around 10 percent of the  total cluster light
(Zibetti  \etal  2005),  have  been tidally  stripped  from  satellite
galaxies.

Formalisms to  describe tidal stripping  in the presence  of dynamical
friction  in a  background  potential have  been  developed by,  among
others, Taylor \& Babul (2001, 2004) and Zentner \& Bullock (2003). As
shown  in   Benson  \etal  (2002),  including  such   a  formalism  in
semi-analytical   models  significantly   reduces  the   abundance  of
satellite galaxies.  Furthermore,  since tidal stripping predominantly
affects satellites  with small pericentric radii,  which are typically
redder (blue  galaxies have only  recently been accreted by  the halo,
and have not yet experienced much dynamical friction), tidal stripping
may also reduce  the red fraction of satellites  and thus help towards
solving the  problem with  the red fractions  of satellites  being too
large.

\subsection{Strangulation}
\label{sec:strangulation}

The second  problem for the  SAM of C06  is that the fraction  of blue
satellite galaxies is much too  low, especially for faint galaxies and
in  low   mass  haloes.    This  suggests  that   `strangulation',  as
incorporated  in the  SAM, is  much too  efficient.  In  virtually all
semi-analytical models,  strangulation is included and  modeled in the
same way: as soon as a  galaxy becomes a satellite galaxy, its hot gas
reservoir  is `stripped'  (i.e.,  the  hot gas  that  belonged to  the
satellite  galaxy is added  to the  hot gas  reservoir of  the central
galaxy).  Consequently, after a delay  time in which the new satellite
galaxy  consumes  (part  of)  its  cold gas,  its  star  formation  is
truncated.  The  results presented here suggest  that this formulation
is too  crude, as it predicts  a blue satellite fraction  that is much
too low\footnote{Note  that the SAM of  C06, as most  other SAMS, does
not even account  for ram pressure stripping, which  will only shorten
the truncation  time, thus  decreasing the blue  satellite fraction.}.
In particular, strangulation is modeled without any explicit halo mass
dependence, which explains why the blue fraction of satellite galaxies
in  the SAM  is  virtually independent  of  halo mass.   In the  SDSS,
however, the  blue satellite  fraction decreases with  increasing halo
mass, suggesting a clear mass scaling of the strangulation efficiency.
The physical mechanisms  thought to be responsible for  the removal of
the  satellite's  hot  gas   reservoir  are  tides  and  ram  pressure
stripping.   The   latter  requires  that   the  parent  halo   has  a
sufficiently dense hot corona.   Since this requirement is more likely
to  be fulfilled  for more  massive haloes,  which in  general  have a
larger   fraction  of  hot   gas,  one   actually  expects   that  the
strangulation efficiency increases with increasing halo mass.  Indeed,
using   numerical   simulations,   Bekki   \etal  (2002)   find   that
strangulation  is  significantly  more  effective  in  massive  galaxy
clusters than  in lower mass groups.  It remains to be  seen whether a
simple scaling along these lines  can bring the red and blue fractions
of satellite galaxies  in SAMs in better agreement  with the SDSS data
presented here.

Interestingly,   in  a  comparison   of  semi-analytical   models  and
hydrodynamical SPH simulations of galaxy formation, Zheng \etal (2005)
have shown  that the  later predicts that  haloes with  $10^{12} \Msun
\lta M  \lta 10^{13}  \Msun$ have a  significantly higher  fraction of
young satellites (similar to  the blue satellites discussed here) than
SAMs, in  much better agreement  with the SDSS results  presented here
(see their Fig.~4).  Similarly, Cattaneo \etal (2006b) find that their
semi-analytic model produces too  few blue satellite galaxies compared
to their SPH simulation. Since these SPH simulations follow the actual
 hydrodynamical  processes   leading  to  strangulation,   this  indeed
suggests that a more realistic  treatment of strangulation in the SAMs
may  solve the  problem with  the colors  of satellite  galaxies.  SPH
simulations such as those described in Zheng \etal (2005) and Cattaneo
\etal (2006b) may prove useful  in calibrating such a new and improved
strangulation model.

\subsection{Dust Extinction and AGN Feedback}
\label{sec:dustagn}

The SAM also has problems with the blue fractions of central galaxies,
which  are too  high, especially  at the  bright and  faint  ends.  In
addition, for a given halo mass, the blue fraction of central galaxies
increases with luminosity, contrary to what is seen in the SDSS. Which
aspect(s)  of  the semi-analytical  model  are  responsible for  these
problems is not  entirely clear. They can indicate  a problem with the
modeling  of dust  extinction, a  problem  with the  treatment of  AGN
feedback, or both.  The former is almost certainly responsible for the
overproduction of  bright and blue centrals.  As  already discussed in
C06,  this  population  of   galaxies  is  reminiscent  of  the  ULIRG
population,  for  which  the  oversimplified  treatment  of  the  dust
extinction is  certainly inadequate: real  starburst are likely  to be
accompanied by  additional extinction which would  make these galaxies
both  fainter and  redder.  This  would  help to  suppress the  strong
increase  of  $f_{\rm blue}$  with  increasing  luminosity, though  it
remains  to be  seen whether  it can  result in  a blue  fraction that
decreases  with increasing luminosity,  as observed.   Furthermore, in
order to suppress  the impact of these starbursting  model galaxies on
the comparison  presented here, we  already removed all  galaxies with
$(g-r)<0$  from  the  SAM.   Despite  this,  however,  the  SAM  still
significantly  overpredicts   the  fraction  of   blue  centrals  with
$^{0.1}M_r - 5\log h \simeq  -23$.  Although improper dust modeling is
likely to  be the cause  for this discrepancy,  it remains to  be seen
whether  this  can  also  explain  the overprediction  (by  $\sim  15$
percent) of the blue fraction of {\it faint} centrals.

Alternatively, the blue fractions  of central galaxies may be modified
by  changing  the  AGN  feedback  description.  As  discussed  in  the
introduction,  it is  still largely  unknown how  AGN impact  on their
surroundings, and thus  feed back on the process  of galaxy formation.
It should therefore not come  as a surprise if the parameterization of
C06 is not  entirely correct or complete.  The  purpose of this paper,
however,  is not  to find  an  improved formulation  of AGN  feedback.
Rather, we have  presented data, in the form  of blue galaxy fractions
as function of  both halo mass and luminosity, which  we believe to be
useful  in  discriminating  between  different  AGN  feedback  models.
Future   SAMs    can   test   and/or    calibrate   their   particular
parameterizations against these data.

\section{Conclusions}
\label{sec:concl}

It  has become  clear that  a  successful reproduction  of the  galaxy
luminosity function, and of the bimodality in the galaxy distribution,
requires a mechanism  that can truncate the star  formation in massive
galaxies at  relatively late epochs. At  the same time,  the fact that
red, passive galaxies preferentially reside in overdense regions, such
as  clusters and  groups  of galaxies,  suggests  that star  formation
truncation  acts  preferentially in  massive  haloes.  Current  models
consider two such truncation  mechanisms: strangulation, which acts on
satellite  galaxies,  and AGN  feedback,  which predominantly  affects
central galaxies.

Typically,  galaxy formation models  are tuned  to reproduce  the {\it
  global}  properties   of  the  galaxy  distribution,   such  as  the
luminosity function and  the total fraction of blue  and red galaxies. 
However,  even  when  two  models  predict  exactly  the  same  global
statistics,  their statistics  as function  of halo  mass may  be very
different.  The latter is clearly more constraining for the model, and
furthermore, holds important information regarding the mass-dependence
of the  various physical mechanisms associated with  galaxy formation. 
In  particular, since  star formation  truncation causes  a  galaxy to
become  red, the  relative fractions  of red  and blue  galaxies  as a
function of  halo mass holds  important clues regarding the  halo mass
dependence of the efficiencies of AGN feedback and strangulation.

To provide a  test-bed for models of galaxy formation,  we have used a
galaxy  group catalogue  extracted from  the  SDSS for  which we  have
computed the  fraction of  blue and red  galaxies as function  of both
galaxy luminosity  and group (halo) mass. To  illustrate the potential
constraining power  of this data  we have compared these  fractions to
those  in the  semi-analytical model  for galaxy  formation  of Croton
\etal  (2006),  which includes  both  `radio  mode'  AGN feedback  and
strangulation.  To allow for a  fair comparison between the SDSS group
catalogue  and  the  SAM,  not influenced  by  potential  inaccuracies
associated with  the group finder  (i.e., interlopers, incompleteness,
errors in assigned group mass),  we have applied the same group finder
over  a mock  redshift survey  constructed from  the SAM.  The $f_{\rm
  blue}(L,M)$  obtained  from this  SAM  group  catalogue  is in  fair
agreement with the true $f_{\rm blue}(L,M)$ obtained directly from the
SAM, indicating  that our group  finder allows a reliable  recovery of
the blue  fraction as  a function of  both galaxy luminosity  and halo
mass.  Although interlopers  and errors in the group  masses may cause
some  errors  in  the  absolute  values of  $f_{\rm  blue}(L,M)$,  the
relative  scalings of $f_{\rm  blue}(L \vert  M)$ and  $f_{\rm blue}(M
\vert L)$ are well recovered.

Although the SAM fits  the overall luminosity function, reproduces the
overall color distribution  of the SDSS galaxies, and  even predicts a
conditional luminosity function whose  shape is in excellent agreement
with  the data,  its  prediction  of $f_{\rm  blue}(L,M)$  is in  poor
agreement with the  SDSS data. In particular, we  have identified four
problems:
\begin{enumerate}
  
\item In massive  haloes the abundance of faint satellite galaxies is
  too high by up to a factor $\sim 2$.
  
\item  The  fraction of  blue  satellite  galaxies  is much  too  low,
  especially for faint galaxies and in low mass haloes.
  
\item The fraction of blue central galaxies is too high, especially at
  the bright  and faint ends.  
  
\item  For a given  halo mass  the blue  fraction of  central galaxies
  increases with luminosity, contrary to what is seen in the SDSS.

\end{enumerate}
The first of these problems is likely  to owe to the fact that the SAM
does  not  model  the tidal  stripping  of  the  stellar mass  of  the
satellite galaxies as they orbit  the parent halo.  As shown in Benson
\etal  (2002),  inclusion of  this  effect  significantly reduces  the
abundance  of satellite galaxies  at a  fixed luminosity.   The second
problem is most  likely due to an oversimplification  of the treatment
of  strangulation. In the  SAM, strangulation  occurs instantaneously,
independent of  halo mass.  However, based  on the SDSS  data, we have
argued that the strangulation efficiency  has to scale with halo mass,
such  that  more massive  haloes  strangulate  their  satellites on  a
shorter time  scale.  The  physical motivation for  such a  scaling is
that  the ram pressure  stripping of  the hot  gas reservoir  of newly
accreted satellites  requires the parent  halo to have  a sufficiently
dense corona of  hot gas.  Since the fraction of  hot gas is typically
an  increasing  function of  halo  mass,  this  may introduce  a  mass
dependence  in  the  strangulation  efficiency as  required.   Indeed,
hydrodynamical  SPH simulations,  which automatically  take  this into
account,   seem  to   predict  blue   satellite  fractions   that  are
significantly higher  than in the semi-analytical  models (Zheng \etal
2005, Cattaneo  \etal 2006b).  Finally,  the third and  fourth problem
listed above,  both of which  concern central galaxies, are  likely to
reflect  shortcomings of the  modeling of  dust extinction  and/or AGN
feedback. 

In summary,  galaxy formation models  are often tested  and calibrated
against  global  properties   of  the  observed  galaxy  distribution.
Recently, with  the inclusion of  AGN feedback, numerous  studies have
claimed success  in reproducing  these global statistics,  even though
very different  formulations for  the various physical  processes have
been used. In order to discriminate between these models more specific
data is  required. In  this paper we  have presented the  fractions of
blue  and red  galaxies  as  function of  luminosity,  halo mass,  and
separately for  central and satellite galaxies.  Clearly  this data is
far  more constraining,  and thus  more challenging,  than  the global
fractions  of  red  and  blue  galaxies,  or  the  overall  luminosity
function.  We have  shown that indeed these data  provide valuable new
insights regarding the  physics of galaxy formation, and  we hope that
they  will provide  a  useful  test-bed for  future  models of  galaxy
formation.

\section*{Acknowledgments}

We thank  Simon D.M. White  for valuable comments.   FvdB acknowledges
useful  discussions with  Eric  Bell, Fabio  Fontanot,  Xi Kang,  Anna
Pasquali and Rachel Somerville.  SW thanks Christian Thalmann for help
with  data handling,  and has  been partially  supported by  the Swiss
National Science  Foundation (SNF).  The work described  in this paper
has made  extensive use  of the Sloan  Digital Sky Survey  (SDSS).  In
particular, we  have used the  New York University  Value-Added Galaxy
Catalogue (http://sdss.physics.nyu.edu/vagc/), and  we are grateful to
Michael Blanton  for his help  with this fantastic data  set.  Funding
for the SDSS has been provided by the Alfred P.  Sloan Foundation, the
Participating  Institutions,   the  National  Aeronautics   and  Space
Administration, the National  Science Foundation, the U.S.  Department
of Energy,  the Japanese Monbukagakusho,  and the Max  Planck Society.
The SDSS Web site is http://www.sdss.org/.  The SDSS is managed by the
Astrophysical   Research  Consortium   (ARC)  for   the  Participating
Institutions.   The Participating Institutions  are The  University of
Chicago,  Fermilab,  the  Institute  for  Advanced  Study,  the  Japan
Participation Group, The Johns Hopkins University, Los Alamos National
Laboratory,  the   Max-Planck-Institute  for  Astronomy   (MPIA),  the
Max-Planck-Institute   for  Astrophysics   (MPA),  New   Mexico  State
University, University of Pittsburgh, Princeton University, the United
States  Naval  Observatory, and  the  University  of Washington.   The
Millennium Run  simulation used in this  paper was carried  out by the
Virgo  Supercomputing  Consortium  at  the  Computing  Centre  of  the
Max-Planck Society in  Garching.  Semi-analytic galaxy catalogues from
the       simulation       are       publicly       available       at
http://www.mpa-garching.mpg.de/galform/agnpaper.

%%%%%%%%%%%%%%%
% Bibliography
%%%%%%%%%%%%%%%

%%%%%%%%%%%%%%%
% Appendices
%%%%%%%%%%%%%%%

\appendix

\section[]{Blue Galaxy Fractions in the SDSS}
\label{sec:AppA}

The following three tables list  the fraction of blue galaxies for all
galaxies  (Table~A1),  for  satellite  galaxies  (Table~A2),  and  for
central  galaxies   (Table~A3),  as  obtained  from   our  SDSS  group
catalogue. Columns correspond to different bins in $\log(M)$, with $M$
in $h^{-1}\Msun$,  as indicated at  the top in square  brackets.  Rows
correspond to different magnitude bins  ($^{0.1}M_{r} - 5 \log h$), as
indicated at  the left in square  brackets. Each entry  lists the blue
fraction  plus,  in  brackets,  the  total number  of  galaxies  (all,
satellite  or central)  in that  bin. As  in  Fig.~\ref{fig:res}, only
entries with at least 50 galaxies are indicated.
\begin{table*}
\label{masslight}
\caption{Blue fractions of all galaxies}
\begin{tabular}{|l|llllll|}
\hline
 & $[12,12.5]$ & $[12.5,13]$ & $[13,13.5]$ & $[13.5,14]$ & $[14,14.5]$ & $[14.5, 15]$\\
\hline\hline
$[-17, -17.5]$ & 0.74 (69)  & -          & -          & -          & -          & -         \\
$[-17.5, -18]$ & 0.68 (108) & 0.61 (114) & 0.56 (80)  & -          & -          & -         \\
$[-18, -18.5]$ & 0.67 (186) & 0.56 (189) & 0.48 (133) & 0.38 (84)  & 0.23 (74)  & -         \\
$[-18.5, -19]$ & 0.60 (308) & 0.54 (343) & 0.44 (265) & 0.28 (213) & 0.22 (96)  & 0.24 (67) \\
$[-19, -19.5]$ & 0.54 (461) & 0.45 (489) & 0.44 (449) & 0.33 (392) & 0.27 (246) & 0.35 (98) \\
$[-19.5, -20]$ & 0.58 (909) & 0.47 (1113)& 0.39 (987) & 0.30 (878) & 0.23 (654) & 0.20 (208)\\
$[-20, -20.5]$ & 0.55 (9366)& 0.44 (1673)& 0.39 (1445)& 0.31 (1172)& 0.23 (771) & 0.17 (296)\\
$[-20.5, -21]$ & 0.50 (7890)& 0.45 (3181)& 0.39 (2556)& 0.32 (1673)& 0.24 (1058)& 0.23 (380)\\
$[-21, -21.5]$ & 0.49 (61)  & 0.42 (4752)& 0.35 (3657)& 0.28 (2164)& 0.26 (1068)& 0.22 (371)\\
$[-21.5, -22]$ & -          & 0.38 (956) & 0.32 (7327)& 0.23 (2231)& 0.21 (1001)& 0.17 (326)\\
$[-22, -22.5]$ & -          &    -       & 0.26 (464) & 0.14 (1181)& 0.11 (476) & 0.08 (143)\\
$[-22.5, -23]$ & -          &    -       &    -       &    -       & 0.06 (108) & -         \\
\hline
\end{tabular}
\medskip

%\begin{minipage}{\hdsize}
%\end{minipage}

\end{table*}
\begin{table*}
\label{masslight}
\caption{Blue fractions of satellite galaxies}
\begin{tabular}{|l|llllll|}
\hline
 & $[12, 12.5]$ & $[12.5, 13]$ & $[13, 13.5]$&$[13.5, 14]$&$[14,14.5]$&$[14.5, 15]$\\
\hline\hline
$[-17, -17.5]$ & 0.74 (69) & -          & -          & -          & -          & -         \\
$[-17.5, -18]$ & 0.68 (108)& 0.61 (114) & 0.56 (80)  & -          & -          & -         \\
$[-18, -18.5]$ & 0.67 (186)& 0.56 (189) & 0.48 (133) & 0.38 (84)  & 0.23 (74)  & -         \\
$[-18.5, -19]$ & 0.60 (308)& 0.54 (343) & 0.44 (265) & 0.28 (213) & 0.22 (96)  & 0.24 (67) \\
$[-19, -19.5]$ & 0.53 (449)& 0.45 (489) & 0.44 (449) & 0.33 (392) & 0.27 (246) & 0.35 (98) \\
$[-19.5, -20]$ & 0.54 (437)& 0.47 (1101)& 0.39 (987) & 0.30 (878) & 0.23 (654) & 0.20 (208)\\
$[-20, -20.5]$ & 0.48 (66) & 0.43 (1235)& 0.39 (1425)& 0.31 (1172)& 0.23 (771) & 0.17 (296)\\
$[-20.5, -21]$ & -         & 0.43 (389) & 0.38 (2005)& 0.32 (1624)& 0.24 (1057)& 0.23 (380)\\
$[-21, -21.5]$ & -         & -          & 0.38 (810) & 0.29 (1568)& 0.26 (1000)& 0.22 (368)\\
$[-21.5, -22]$ & -         & -          & -          & 0.26 (408) & 0.23 (621) & 0.18 (283)\\
\hline
\end{tabular}
\medskip

%\begin{minipage}{\hdsize}
%\end{minipage}

\end{table*}
\begin{table*}
\label{masslight}
\caption{Blue fractions of central galaxies}
\begin{tabular}{|l|llllll|}
\hline
 & $[12, 12.5]$ & $[12.5, 13]$ & $[13, 13.5]$&$[13.5, 14]$&$[14,14.5]$&$[14.5, 15]$\\
\hline\hline
$[-19.5, -20]$ & 0.61 (472) & -          & -          & -         & -         &  -       \\
$[-20, -20.5]$ & 0.55 (9300)& 0.47 (438) & -          & -         & -         & -        \\
$[-20.5, -21]$ & 0.51 (7890)& 0.46 (2792)& 0.42 (551) & -         & -         & -        \\
$[-21, -21.5]$ & 0.49 (61)  & 0.42 (4752)& 0.34 (2847)& 0.26(596) & 0.28 (68) & -        \\
$[-21.5, -22]$ & -          & 0.38 (956) & 0.32 (7324)& 0.22(1823)& 0.18 (380)& -        \\
$[-22, -22.5]$ & -          & -          & 0.26 (464) & 0.14(1178)& 0.11 (444)& 0.07(95) \\
$[-22.5, -23]$ & -          & -          &    -       &  -        & 0.06 (108)& -        \\
\hline
\end{tabular}
\medskip

%\begin{minipage}{\hdsize}
%\end{minipage}

\end{table*}

\end{document}